\documentclass[fleqn,usenatbib,referee]{mnras}

\usepackage{subfig}
\usepackage{float}
\usepackage{footnote}
\usepackage{color,soul}
\renewcommand{\hl}[1]{{\color{red}#1}}
\makesavenoteenv{tabular}

\usepackage[T1]{fontenc}
\usepackage{ae,aecompl}
\usepackage{gensymb}

\usepackage{graphicx}	
\usepackage{amsmath}	
\usepackage{amssymb}	
\usepackage{multirow}
\usepackage{newtxtext,newtxmath}

\title[2S 1417$-$624]{Timing and spectral analysis of 2S 1417$-$624 during its 2018 outburst}

\author[Serim et al.]
{M. M. Serim$^{1,2,3}$, \"{O}. C. \"{O}z\"{u}do\u{g}ru$^{1}$\thanks{E-mail: ozudogru@astroa.physics.metu.edu.tr (\"{O}C\"{O}); muhammed@astroa.physics.metu.edu.tr (MMS); inam@baskent.edu.tr (S\c{C}\.{I})}, \c{C}. K. D\"{o}nmez$^{1}$, 
\c{S}. \c{S}ahiner$^{4}$,
D. Serim$^{1}$, A. Baykal$^{1}$
\newauthor and S. \c{C}. \.{I}nam$^{5}$
\\
$^{1}$Department of Physics, Middle East Technical University, 06800 Ankara, Turkey\\
$^{2}$Institut f\"{u}r Astronomie und Astrophysik, Sand 1, 72076 T\"{u}bingen, Germany\\
$^{3}$Department of Electrical and Electronics Engineering, At{\i}l{\i}m University, 06830 Ankara, Turkey\\
$^{4}$Department of Electronics and Communication Engineering, 
Beykent University, 34398 \.{I}stanbul, Turkey \\
$^{5}$Department of Electrical and Electronics Engineering, 
Ba\c{s}kent University, 06790 Ankara, Turkey
}

\date{Accepted XXX. Received YYY; in original form ZZZ}

\pubyear{2021}

\begin{document}
\label{firstpage}
\pagerange{\pageref{firstpage}--\pageref{lastpage}}
\maketitle


\begin{abstract}

We investigate timing and spectral characteristics of the transient X-ray pulsar 2S 1417$-$624 during its 2018 outburst with \emph{NICER} follow up observations.
We describe the spectra with high-energy cut-off and partial covering fraction absorption (PCFA) model and present flux-dependent spectral changes of the source during the 2018 outburst. Utilizing the correlation-mode switching of the spectral model parameters, we confirm the previously reported sub-critical to critical regime transitions and we argue that secondary transition from the gas-dominated to the radiation pressure-dominated disc do not lead to significant spectral changes below 12 keV. Using the existing accretion theories, we model the spin frequency evolution of 2S 1417$-$624 and investigate the noise processes of a transient X-ray pulsar for the first time using both polynomial and luminosity-dependent models for the spin frequency evolution. For the first model, the power density spectrum of the torque fluctuations indicate that the source exhibits red noise component ($\Gamma \sim -2$) within the timescales of outburst duration which is typical for disc-fed systems. On the other hand, the noise spectrum tends to be white on longer timescales with high timing noise level that indicates an ongoing accretion process in between outburst episodes. For the second model, most of the red noise component is eliminated and the noise spectrum is found to be consistent with a white noise structure observed in wind-fed systems.

\end{abstract}


\begin{keywords}
stars: neutron - pulsars: individual: 2S 1417$-$624 - accretion, accretion discs
\end{keywords}



\section{Introduction}
2S 1417$-$624 is a transient Be X-ray binary (BeXB), discovered by Small Astronomy Satellite 3 (\emph{SAS-3}) in 1978 \citep{Apparao1980}. Shortly after its discovery, 17.64 second (0.5669 Hz) pulsations were revealed and the source was observed to spin-up at a rate of {$\sim$(3-6)$\times$10$^{-11}$ Hz s$^{-1}$} \citep{Kelley1981}. An optical companion of B1 V\textit{e} type with $\sim$16.9 mag was detected, and its distance was inferred to be between 1.4 to 11.1 kpc \citep{Grindlay1984}. Recent distance estimations provided by \emph{Gaia} yielded a result of $9.9^{+3.1}_{-2.4}$ kpc \citep{BailerJones2018}; however, \citet{Ji2020} discussed the possibility of a 20 kpc distance by cross-resolving the critical luminosity and torque equations on the distance-magnetic field diagram. The orbital period of the binary and the eccentricity of the orbit are determined to be $\sim$42.19 days and 0.446, respectively \citep{Finger1996,Inam2004}, which is further refined by the \emph{Fermi}/GBM team\footnote{\label{gbmfn}\url{https://gammaray.nsstc.nasa.gov/gbm/science/pulsars/lightcurves/2s1417.html}}.

The second giant outburst of the source, after its discovery was observed in 1994, and right after it is followed by several smaller Type-I outbursts, all of which were observed by Burst and Transient Source Experiment (\emph{BATSE}) onboard \textit{Compton} Gamma Ray Observatory \citep{Finger1996}. The third outburst during 1999--2000 was meticulously kept under observation with Rossi X-ray Timing Explorer (\emph{RXTE}) apart from \emph{BATSE}. The correlation between pulse fraction and source flux, and a double-peaked pulse profile during the outburst were reported \citep{Inam2004}. In 2009, the fourth giant outburst was detected by \emph{Fermi} Gamma-ray Burst Monitor (GBM), and Burst Alert Telescope (BAT) on-board Neil Gehrels \textit{Swift} Observatory \citep{Beklen2009, Krimm2009}, and also observed by \emph{RXTE} subsequently \citep{Gupta2018}. This time, along with a double-peaked pulse profile, triple-peaked pulse profiles were also observed around the maximum luminosity \citep{Gupta2018}. Furthermore, an anti-correlation between pulse fraction and the source flux was reported, which was argued to be the result of the change of the accretion geometry from pencil beam to a combination of pencil and fan beam \citep{Gupta2018}. The fifth outburst in 2018, reaching to $\sim$350 mCrab intensity, was the brightest giant outburst so far. It was observed that the pulse profiles are evolving into a more complicated shape with increasing luminosity, becoming quadruple-peaked at maximum luminosity, which was not observed previously during the less luminous giant outbursts \citep{Ji2020}. Pulse fraction was found to be correlated with luminosity in the hard band (25--100 keV) using \textit{Insight-HXMT} (Hard X-ray Modulation Telescope: \citeauthor{Zhang2014} \citeyear{Zhang2014}) data. During these outbursts, the spin-up rate was found to be scaling up with the increasing pulsed flux, therefore 2S 1417$-$624 is a good candidate for testing accretion models. In 2021 January, 2S 1417$-$624 entered another outburst episode, which has faded out since then \citep{Hazra2021, Knies2021}.

The X-ray spectra of 2S 1417$-$624 are generally modeled with an absorbed power-law with a high-energy cut-off, plus an iron emission line at $\sim$6.4--6.8 keV \citep{Finger1996, Inam2004}. No cyclotron emission features have been observed from 2S 1417$-$624 in the energy range of 0.9--79 keV \citep{Gupta2019}. 2S 1417$-$624 was also observed in its quiescent state with \emph{Chandra} in 2013 May with 4.7 $\times$ 10$^{-13}$ erg s$^{-1}$ cm$^{-2}$ unabsorbed flux \citep{Tsygankov2017}. No pulsations from the source were detected, and it is argued that the quiescent spectrum may be described by either a power-law model or a thermal component with a temperature of $\sim$1.5 keV.

Several studies investigated the timing noise of accretion-powered X-ray pulsars in terms of statistical correlations or more source-specific time-dependent noise strength variations (e.g. \cite{Baykal1993,Bildsten1997, DAlessandro1995,Baykal1997}). Timing noise can originate from internal and external torques acting on the neutron star. In the case of accretion-powered X-ray pulsars, \citet{Baykal1993} suggested that the timing noise is connected to mass accretion by revealing a statistical correlation with X-ray luminosity. \citet{Bildsten1997} examined the power density spectra of the torque fluctuations of 8 persistent systems to probe the nature of the accretion in these systems. They reported white torque-noise structure in known wind-fed systems, consistent with the expectations from the simulations of timing noise in such systems \citep{Fryxell1988}, since their spin frequency evolution can be characterized as random shots of independent torque fluctuations.
On the other hand, disk-fed systems should exhibit a long-term correlation of accretion torques which can lead to red-noise components in the power density spectra of the torque fluctuations. So far, timing noise studies focus only on persistent and isolated sources. In this paper, we present a study of the torque fluctuations of a transient system using luminosity-dependent spin frequency modeling for the first time.
 
\section{Observations and Data Reduction}
\label{data_reduction}
In this study, we use target of opportunity observations of 2S 1417$-$624 during its outburst in 2018, which were taken by the Neutron Star Inner Composition Explorer (\emph{NICER}) X-Ray Timing Instrument (XTI) stationed at the International Space Station and operated by NASA \citep{Gendreau2016}. \emph{NICER}/XTI detects soft X-rays in energy range 0.2--12 keV with high timing resolution of about 100 ns. It has 56 co-aligned X-ray concentrators that focus X-rays to Focal Plane Modules composed of silicon drift detectors \citep{Prigozhin2012}. Out of 56 detectors, four are inactive since launch and two detectors (14, 34) are reported to have increased instrumental noise\footnote{\url{https://heasarc.gsfc.nasa.gov/docs/nicer/data_analysis/nicer_analysis_tips.html}}, occasionally. Therefore these detectors are excluded from our analysis.

\emph{NICER} master archive catalog contains 85 observations of 2S 1417$-$624 between 2018 April 1–September 3. 17 of the observations have zero or very low exposures ($< 100$ s) after data screening therefore we use 68 observations which add up to a total exposure of 82.9 ks  {\hl(see Table \ref{obslist})}. We implement the most recent calibration files\footnote{\url{https://heasarc.gsfc.nasa.gov/docs/heasarc/caldb/data/nicer/xti/index.html}} (version of July 22, 2020) available at the time of the analysis. We use gain calibration file version 6 in our analyis. In the data reduction and calibration process, we used \textsc{nicerdas} version 7a which is issued within \textsc{heasoft} v6.28\footnote{\url{https://heasarc.gsfc.nasa.gov/docs/software/lheasoft/download.html}}. The clean events and mkf files are reproduced by using standard level 2 data processing steps (\texttt{nicerl2} tool) with the default parameters recommended by the \emph{NICER} team. These include screening off times when the pointing offset is greater than 0.015\degree, the distance to the bright earth limb is greater than 30\degree, the distance to the dark earth limb is greater than 15\degree and also times within the South Atlantic Anomaly. 

\begin{table*}
  \caption{List of NICER observations used in this work. }
  \label{obslist}
  \center{\renewcommand{\arraystretch}{1.0}\begin{tabular}{rclrclrcl}
  \hline  
Obs. ID	&	Time (MJD)	&	Exposure (s)	&	Obs. ID	&	Time (MJD)	&	Exposure (s)	&	Obs. ID	&	Time (MJD)	&	Exposure (s)	\\
    \hline 
1200130101	&	58209.51	&	563	&	1200130133	&	58260.00	&	1890	&	1200130156	&	58297.42	&	1718	\\
1200130104	&	58214.49	&	986	&	1200130134	&	58261.16	&	373	&	1200130157	&	58298.45	&	894	\\
1200130105	&	58215.01	&	399	&	1200130135	&	58262.07	&	1068	&	1200130158	&	58299.67	&	497	\\
1200130106	&	58219.38	&	199	&	1200130136	&	58263.23	&	1034	&	1200130160	&	58301.47	&	655	\\
1200130107	&	58222.00	&	592	&	1200130137	&	58264.07	&	1223	&	1200130165	&	58308.24	&	1594	\\
1200130108	&	58223.03	&	1063	&	1200130138	&	58268.16	&	517	&	1200130166	&	58310.18	&	879	\\
1200130109	&	58224.06	&	120	&	1200130139	&	58269.38	&	1209	&	1200130167	&	58311.33	&	1140	\\
1200130110	&	58225.02	&	268	&	1200130140	&	58271.18	&	1593	&	1200130168	&	58312.43	&	1112	\\
1200130114	&	58233.12	&	3938	&	1200130141	&	58272.53	&	2692	&	1200130169	&	58317.56	&	707	\\
1200130115	&	58236.34	&	1520	&	1200130142	&	58273.37	&	637	&	1200130171	&	58321.55	&	901	\\
1200130116	&	58237.56	&	1334	&	1200130143	&	58274.34	&	1044	&	1200130172	&	58322.77	&	307	\\
1200130117	&	58238.07	&	1808	&	1200130144	&	58275.04	&	1570	&	1200130173	&	58323.68	&	718	\\
1200130118	&	58239.30	&	2231	&	1200130145	&	58276.01	&	2366	&	1200130174	&	58324.64	&	720	\\
1200130119	&	58240.79	&	1219	&	1200130146	&	58278.65	&	963	&	1200130175	&	58326.12	&	1292	\\
1200130120	&	58241.22	&	3743	&	1200130147	&	58279.03	&	920	&	1200130176	&	58327.73	&	114	\\
1200130122	&	58244.12	&	1222	&	1200130148	&	58280.06	&	1815	&	1200130177	&	58328.82	&	822	\\
1200130123	&	58245.08	&	1133	&	1200130149	&	58282.19	&	685	&	1200130178	&	58329.79	&	534	\\
1200130124	&	58246.05	&	3426	&	1200130150	&	58283.99	&	888	&	1200130179	&	58330.89	&	596	\\
1200130126	&	58248.11	&	3603	&	1200130151	&	58289.06	&	1373	&	1200130181	&	58338.77	&	1488	\\
1200130127	&	58249.92	&	562	&	1200130152	&	58290.67	&	1865	&	1200130183	&	58348.23	&	590	\\
1200130128	&	58250.05	&	1800	&	1200130153	&	58292.08	&	1164	&	1200130184	&	58354.34	&	724	\\
1200130129	&	58251.15	&	2780	&	1200130154	&	58293.75	&	476	&	1200130185	&	58364.75	&	540	\\
1200130130	&	58252.05	&	779	&	1200130155	&	58296.13	&	1669	&		&		&	\\    
\hline
  \end{tabular}}  
  
  \end{table*}  

The source and background spectra of each observation are extracted with the background estimator tool \texttt{nibackgen3C50} version 6 using 2020 gain calibration, one of the recommended tools by the \emph{NICER} team\footnote{\url{https://heasarc.gsfc.nasa.gov/docs/nicer/tools/nicer_bkg_est_tools.html}}. During the spectral analysis, data below 0.8 keV are excluded due to high background count rates comparable to source spectrum. Generated spectra are rebinned to have at least 30 counts per bin via \texttt{grppha} tool. Data below 0.8 keV are excluded during spectral analysis, because of calibration issues. Before applying version ‘20170601v004’ of the ancillary response file and version ‘20170601v002’ of the redistribution matrix file, we adjust them for the exclusion of noisy detectors as recommended\footnote{\url{https://heasarc.gsfc.nasa.gov/docs/nicer/analysis_threads/arf-rmf-precomputed/}}. Spectra are modelled with \textsc{xspec} v12.11.1. For the timing analysis, event time series are corrected for the barycenter of the Solar system using \texttt{barycorr} tool with JPL ephemeris DE430. Lightcurves are extracted in \textsc{XSELECT} with a time resolution of 0.125 s. 

\section{Spectral Analysis}

We investigate 0.8--12 keV spectra of 68 \emph{NICER} observations of 2S 1417$-$624, covering a time span of $\sim 150$ days during its outburst in 2018. The spectra of 2S 1417$-$624 can be described by photoelectric absorption and power-law with a high-energy cut-off model, and also an additional Gaussian component to characterize
the iron fluorescent line around $\sim$6.4--6.7 keV \citep{Finger1996, Inam2004}. Even though stated model is sufficient to explain the spectral continuum at lower flux levels, the spectra at higher flux levels become more complicated and they require additional components to properly describe the data. For example, \cite{Gupta2019} investigated the spectra (0.9--79 keV) of 2S 1417$-$624 with \emph{NuSTAR} and \emph{Swift}/XRT which almost simultaneously observed the source at the peak of the outburst in 2018. They tested several alternative models to represent the data, that include addition of a thermal component (with kT $\sim 0.94$ keV) or a partial covering fraction absorption (PCFA). 

\subsection{Time-averaged spectra}

\begin{table*}
  \caption{List of spectral parameters for the peak of the outburst obtained via analysis of merged observations between MJD 58233.1 -- 58252.1. Model 1 is \texttt{PHABS*PCFABS*(POW*HIGH+GAU)} and model 2 is \texttt{PHABS*(POW*HIGH+BBODYRAD+GAU)}. Uncertainties are given for 90\% confidence interval. }
  \label{spepar}
  \center{\renewcommand{\arraystretch}{1.2}\begin{tabular}{llcc}
  \hline
Component & Parameters	&	Model 1	&	Model 2	\\
	\hline
phabs & $n_{H_1}$ (10$^{22}$ cm$^{-2}$)	&	1.12 (fixed)	&	$1.07 \pm 0.01$	\\
pcfabs & $n_{H_2}$ (10$^{22}$ cm$^{-2}$)	&	55.0$^{+5.3}_{-4.8}$	&	--	\\
 & Covering fraction	&	$0.24 \pm 0.02$	&	--	\\
powerlaw & $\Gamma$	&	$0.738 \pm 0.004$ 	&	$0.43 \pm 0.02$	\\
 & norm (10$^{-2}$)	&	$8.5 \pm 0.2$	&	$4.1 \pm 0.2$	\\
highecut & $E_{\text{cut}}$ (keV)	&	8.3$^{+0.2}_{-0.3}$	&	$6.7  \pm 0.1$	\\
 & $E_{\text{fold}}$ (keV)	&	19.0$^{+3.8}_{-2.8}$	&	19.5$^{+2.0}_{-1.8}$	\\
bbodyrad & kT (keV)	&	--	&	0.56$^{+0.01}_{-0.02}$	\\
 & R$^{(a)}$ (km)	&	--	&	8.0$^{+0.3}_{-0.2}$	\\
gaussian & Line energy (keV)	&	$6.40 \pm 0.02$	&	$6.40 \pm 0.01$	\\
 & $\sigma$ (keV)	&	$0.07 \pm 0.02$	&	$0.09 \pm 0.02$	\\
 & norm  (10$^{-4}$)	&	$6.6 \pm 0.8$ 	&	$6.5 \pm 0.8$ 	\\
 & Equivalent width (eV)	&	30$^{+4}_{-3}$	&	36$^{+4}_{-3}$	\\
 & & \\
 & Blackbody Flux$^{(b)}$ &	--	&	0.63$^{+0.06}_{-0.05}$	\\ 
 & Unabsorbed Flux$^{(b)}$ 	&	$23.5 \pm 0.5$	&	$19.36 \pm 0.07$	\\
\hline
 & $\chi^{2}_{\text{Red}}$ ($\textit{d.o.f.}$)	&	0.98 (1104)	&	0.88 (1103)	\\
\hline
  \end{tabular}}
\begin{flushleft}
$\bf{Notes.}$ 

$^{(a)}$ Blackbody emitting radius is calculated assuming a distance of 9.9 kpc.

$^{(b)}$ 0.8 - 12 keV flux values are given in units of 10$^{-10}$ erg cm$^{-2}$ s$^{-1}$

\end{flushleft}  
  
  \end{table*}
  
In order to observe the features in the \emph{NICER} spectrum during the peak of the 2018 outburst, we merge the event files between MJD 58233.1 -- 58252.1 with the tool \texttt{nimpumerge} and create a time-averaged spectrum for the peak of the outburst. The total exposure is 31.1 ks during this time interval. We add a systematic error of 1\% during the analysis of the merged spectrum during the peak of the outburst. We first model this spectrum with a model composed of an absorbed powerlaw with a high-energy cutoff and a Gaussian component at 6.4 keV for iron K$\alpha$ emission line. However this fit is inadequate with a reduced $\chi^2$ of 1.66. Therefore we try the model that includes a thermal component represented by a blackbody (BB) emission. The best-fit parameters of this model are given in Table \ref{spepar} as Model 2. Addition of the BB component decreases the reduced $\chi^2$ to 0.88 which implies that this model overfits the data. Furthermore, BB temperature is estimated to be 0.56 keV which is less than the estimation of \cite{Gupta2019}. The BB emission radius calculated from the normalization of \texttt{BBODYRAD} model assuming a distance of 9.9 kpc (\emph{GAIA} value) is about 8.0 km. Since we aim to search for a time evolution of spectral parameters, we also try this model for single observations and notice that the BB component cannot be resolved in all observations. Moreover the photon index evolution cannot be traced correctly because lower photon index values are found when BB is added to the model only in some of the observations.

\begin{figure}
	\includegraphics[width=0.9\columnwidth]{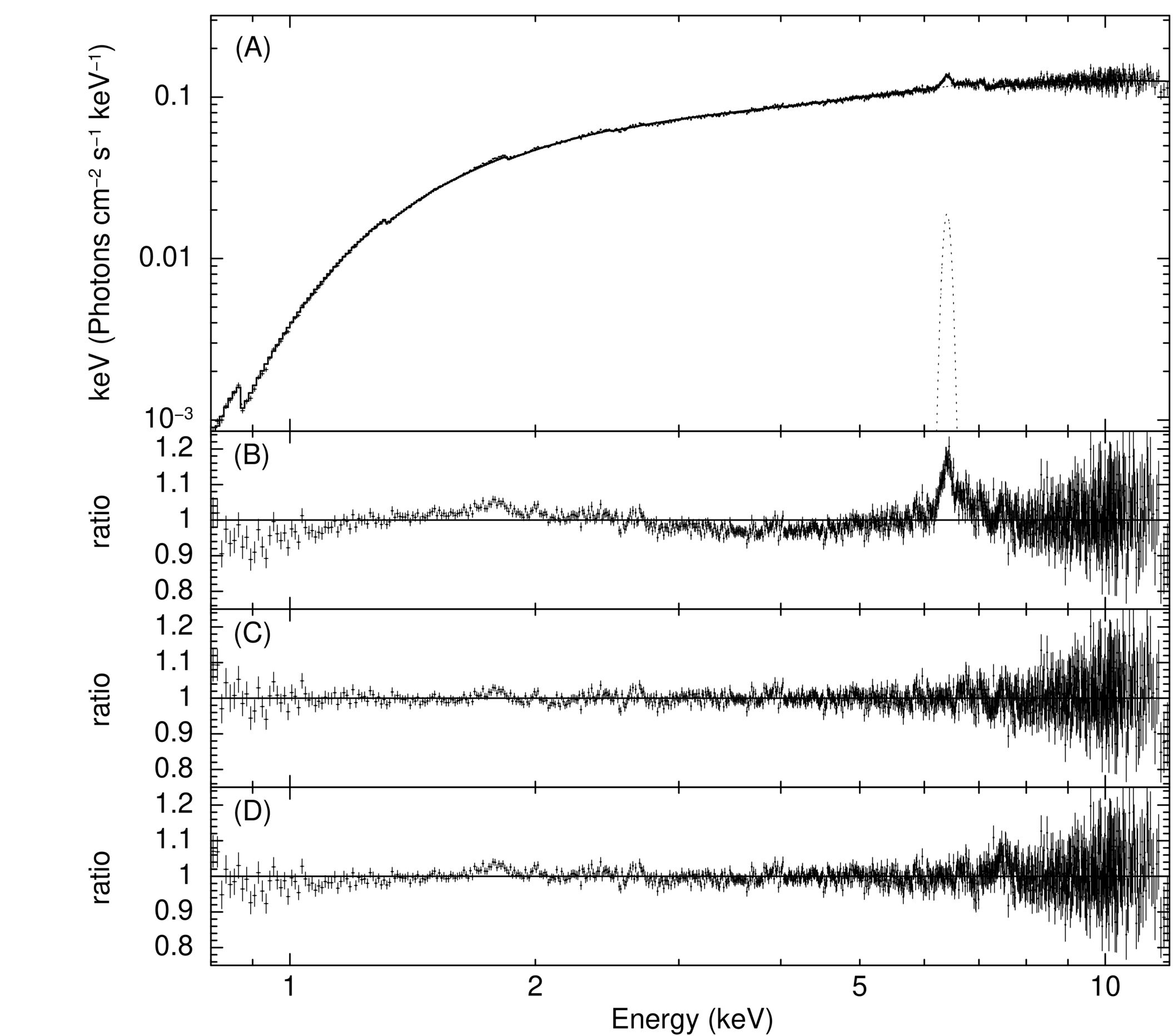}
	\caption{0.8-12 keV energy spectrum of 2S 1417$-$624 for the peak of the outburst obtained by merging observations between MJD 58233.1 -- 58252.1. Panel A shows the data and the best fit (solid line) for \texttt{PHABS*PCFABS*(POW*HIGH+GAU)} model. Dotted curve indicates the gaussian component fitted at 6.4 keV for the Fe K$\alpha$ emission. Panel D demonstrates the ratio of data to this model, whereas panel B is the ratio for \texttt{PHABS*POW*HIGH} model and panel C is the ratio for \texttt{PHABS*(POW*HIGH+BBODYRAD+GAU)} model.} 
	\label{fig_spectrum}
\end{figure}

Next we model the merged spectrum during the peak of the outburst using a model composed of an absorbed powerlaw with a high-energy cutoff plus a partial covering absorber. Fixing the hydrogen column density of the photoelectric absorption model ($n_{H_1}$)  to 1.12 $\times$ 10$^{22}$ atoms cm$^{-2}$, that is the galactic average\footnote{\url{https://heasarc.gsfc.nasa.gov/cgi-bin/Tools/xraybg/xraybg.pl}} in the direction of 2S 1417$-$624 , the hydrogen column density of the partial covering absorber ($n_{H_2}$) would characterize the absorbing material in the vicinity of the source. This model gives an acceptable fit with a reduced $\chi^2$ of 0.98. The best-fit parameters are given in Table \ref{spepar} as Model 1 and the energy spectrum is shown in Fig. \ref{fig_spectrum}. It has been observed that in some cases spectra of Be/X-ray binaries require an additional absorber to establish an adequate continuum model \citep{Naik2011,Epili2017,Jaisawal2016}.  We use this model for single observations and perceive that it is successful for all of the observations, therefore suitable to investigate time-resolved spectra.  

\subsection{Time-resolved spectra}
Single observation exposures are varying from 3.9 ks to 0.1 ks and are generally below 1 ks (see Table \ref{obslist}). 
During the spectral analysis of single observations high-energy cutoff is excluded from Model 1 and no systematic error is added.
Since we confront with weak iron line in low exposure and low flux observations, we fix the line energy and sigma to their best fit values found from the time-averaged spectrum during the peak of the outburst. Therefore, we trace the variation of the covering fraction and column density of the PCFA model, photon index and unabsorbed flux.

\begin{figure*}
	\includegraphics[width=0.49\columnwidth]{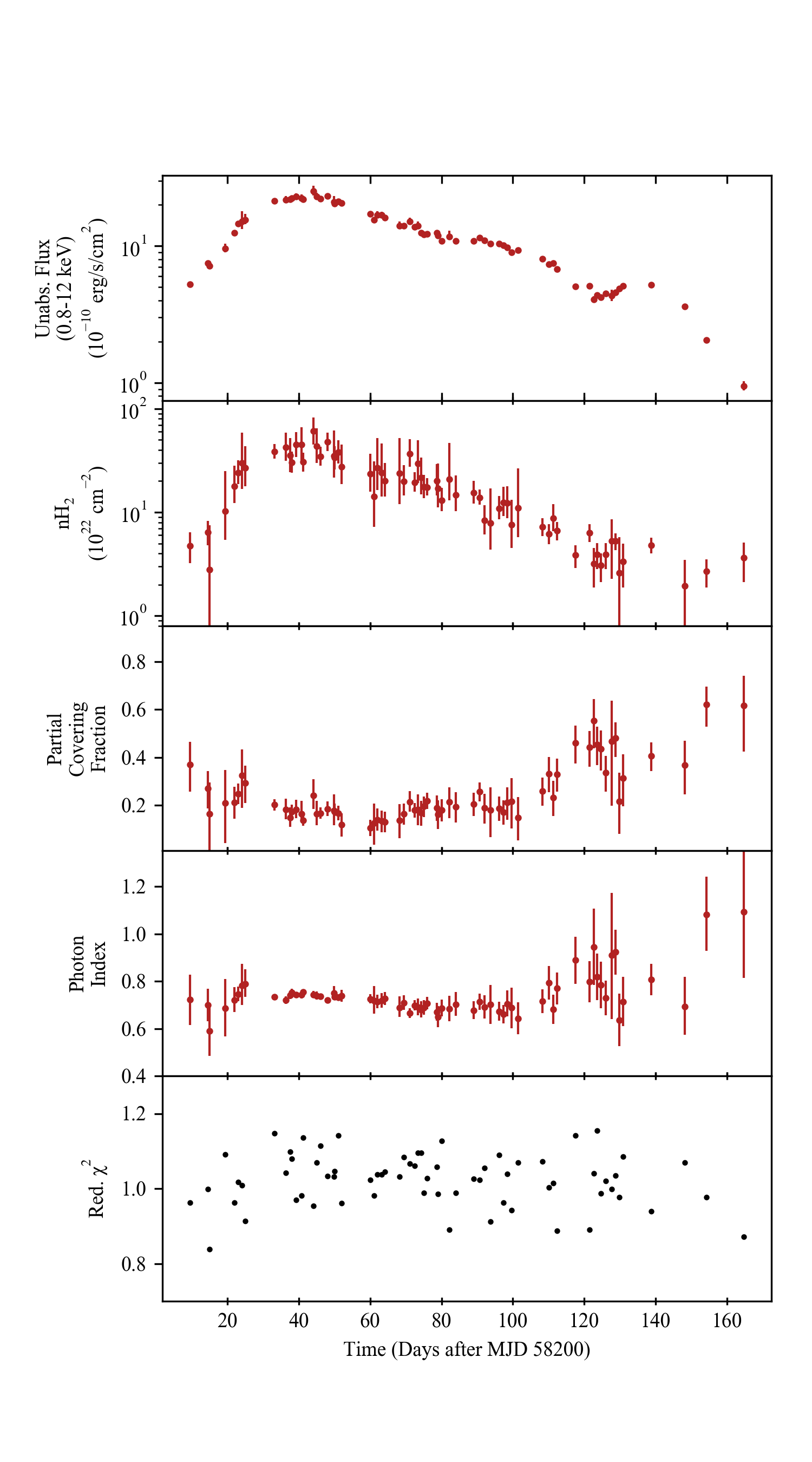}
	\includegraphics[width=0.49\columnwidth]{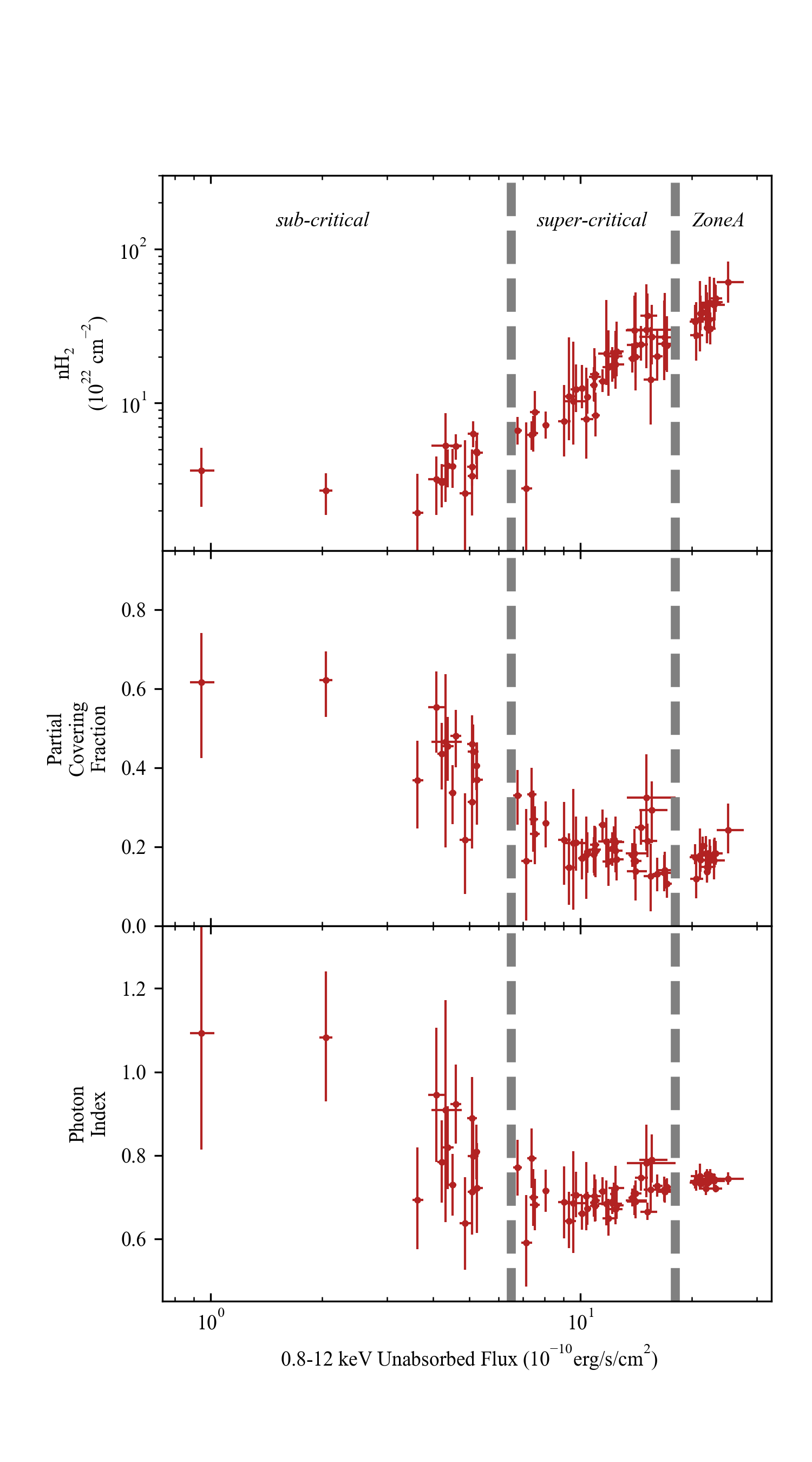}
	\caption{Left: The time evolution of the spectral parameters of 2S 1417$-$624 during its outburst in 2018. Right: The flux dependence of the model parameters. The flux levels associated with the accretion regime transitions $L_{\rm{crit}}$ and $L_{\rm{ZoneA}}$ (see the text) are indicated with vertical dashed lines.Uncertainties are given for 90\% confidence interval.}
	\label{fig_specevo_time}
\end{figure*}

The results obtained from the spectral modeling of single observations are illustrated in Fig. \ref{fig_specevo_time}. Overall, the model provides a statistically adequate description of individual spectra, with $\chi^2 < 1.1$ in almost all cases. Unabsorbed flux measurements in 0.8--12 keV energy band (Fig. \ref{fig_specevo_time}, top left panel) exhibits a similar profile with the one observed in 15--50 keV \emph{Swift}/BAT count rate. Time evolution of the PCFA model parameters and photon index show sporadic correlations and anti-correlations with flux. Hence, we further examine flux dependence of the parameters (Fig. \ref{fig_specevo_time}, right). Furthermore, we also illustrate the corresponding flux levels of accretion regime transitions, $L_{\rm{crit}}$ and $L_{\rm{ZoneA}}$ as reported by \citet{Ji2020}. The flux levels that correspond to $L_{\rm{crit}}$ and $L_{\rm{ZoneA}}$ are interpolated by using the \emph{NICER} flux measurements at the dates close to the accretion regime turnovers reported by \citet{Ji2020} (MJD $\sim$58308 for $L_{\rm{crit}}$  and MJD $\sim$58260 for  $L_{\rm{ZoneA}}$). Both partial covering fraction and the photon index are in anti-correlation with the flux up to $L_{\rm{crit}}$; however, their correlations change above this flux level. Second transition level, which corresponds to $L_{\rm{ZoneA}}$, do not reflect significant changes in the flux dependence of spectral parameters. When the source luminosity is below $L_{\textrm{crit}}$, the covering fraction and photon index decreases from $\sim$0.6 to $\sim$0.2 and $\sim$1.1 to $\sim$0.6 respectively, as the source luminosity increases (Pearson correlation coefficients, $p$, $-0.75$ and $-0.71$, respectively). Even though the spectral softening is not very prominent at super-critical regime (photon indices remain in the range from $\sim$0.6 to $\sim$0.7), the flux measurements still seem to be moderately correlated with the photon index (with p $\sim$ 0.7 for super-critical and zoneA regimes combined). Beyond $L_{\rm{crit}}$, the covering fraction becomes uncorrelated and resides in the range from $\sim$0.1 to $\sim$0.3. On the other hand, the column density of the PCFA remains more or less constant below $L_{\rm{crit}}$ and becomes strongly correlated with the flux above the critical level ($p\sim 0.93$ for supercritical and zoneA regimes combined), reaching up to $\sim 6\times 10^{23}$ atoms cm$^{-2}$ when the source is brightest.

\section{timing analysis}

\subsection{Pulse Frequency Measurements}
2S 1417$-$624 is regularly monitored with \emph{Fermi}/GBM \citep{Meegan2009} during its outburst in 2018 and the GBM Accreting Pulsars Program (GAPP) team shares out their timing analysis and pulsed flux measurement results through their web page\footnote{\url{https://gammaray.nsstc.nasa.gov/gbm/science/pulsars/lightcurves/2s1417.html}}. The spin frequency measurements of the GAPP team are provided as orbitally corrected for Doppler delays with a slightly improved version of the orbital solution reported by \cite{Inam2004}. For our analysis, we make use of the orbitally-corrected frequency history generated by the GAPP team.
\begin{figure}
	\includegraphics[width=\columnwidth]{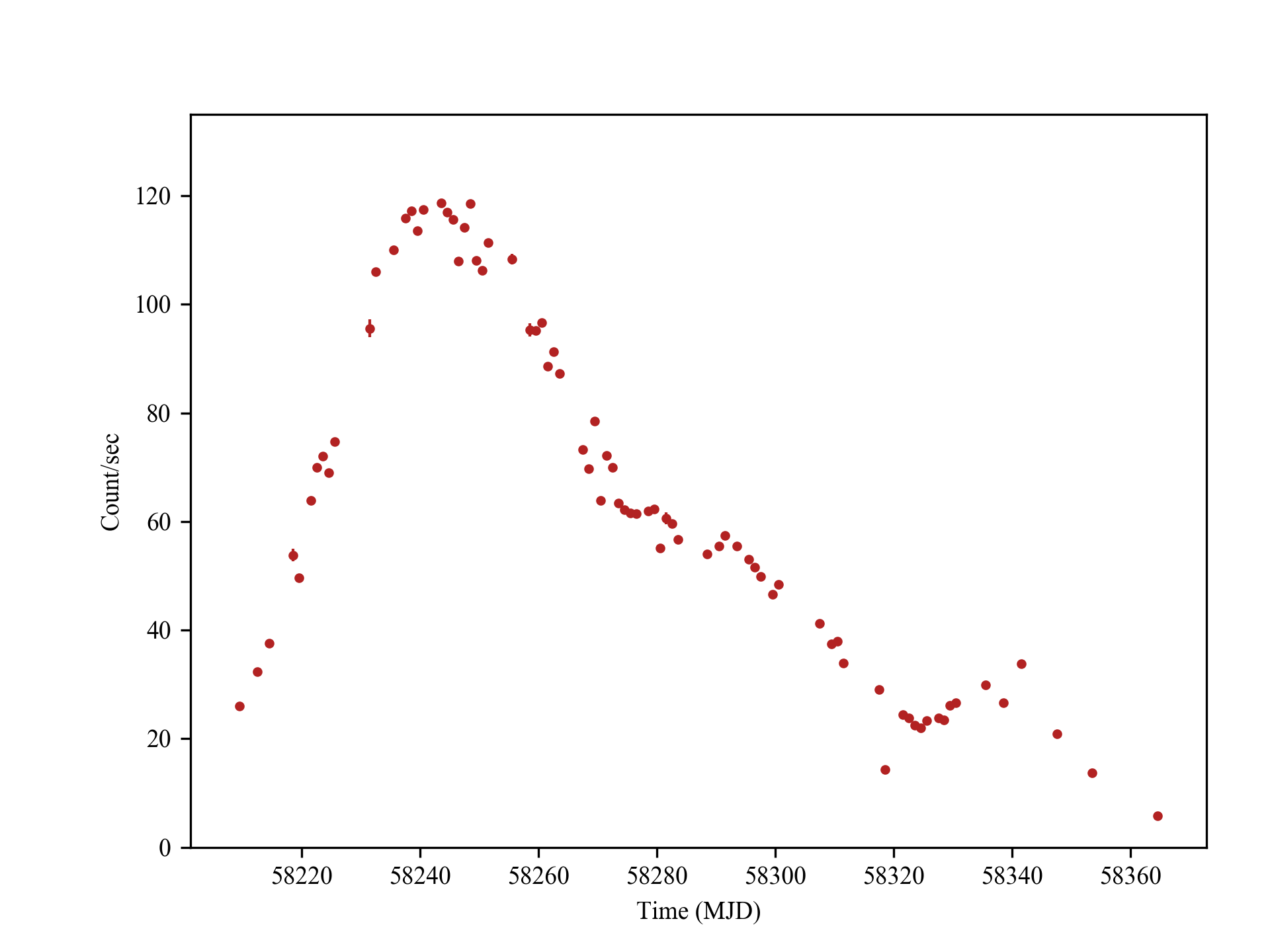}
	\caption{One-day binned lightcurve of 2S 1417$-$624 obtained from \emph{NICER}/XTI observations in 0.8-12 keV energy band}
	\label{fig_lightcurvel}
\end{figure}
For the \emph{NICER}/XTI observations, we first apply orbital correction to the lightcurve (Fig. \ref{fig_lightcurvel}) with the orbital solution described by the GAPP team. Then, we seek for local timing solutions (covering a time span of $\sim$70 days) in \emph{Fermi}/GBM frequency history by fitting 6$^{th}$ order polynomials within each segment. The timing parameters of each segment is listed in Table \ref{timing_params}. Afterwards, we fold the \emph{NICER} observations on the local timing solutions to obtain the pulse profiles (with 20 phase bins) for each observation. When the observation sequence has a gap less than half-day, they are combined to generate one single pulse profile. The pulse profiles are described analytically in the form of harmonic functions \citep{Deeter1982} and cross-correlated to calculate pulse arrival times (TOAs). For each local segment, the cross-correlation analysis is performed by taking the luminosity-dependent pulse profile variability \citep{Ji2020} into account (i.e. we solely employ the pulses with same profiles in each segment). 
However, complex luminosity-dependent pulse shape and high spin-up rates during the outburst make it hard to obtain a phase coherent timing solution for long time spans. Instead, we use the solutions obtained by fitting the \emph{Fermi}/GBM measurements and we calculate piecewise derivative of the pulse TOAs by fitting a linear model to each sequential TOA pair. By utilizing the slope of each fit, $ \delta\nu = \delta\phi / (t_2-t_1)$, where $t_1$ and $t_2$ are the time of the first and second pulse of TOA pair, we convert them into spin frequency measurements at the mid-time of each interval (see \cite{Cerri2019} for applications). Uncertainty in spin frequency measurements are derived from $1 \sigma$ error ranges of slopes obtained during the fit. The spin frequency history generated from \emph{NICER} observations are illustrated in Fig. \ref{fig_freq_model} (upper panel) along with the spin frequency history provided by the GAPP team.

\begin{table*}
\centering
\caption{Table of timing parameters obtained from modeling the spin frequency measurements provided by the GAPP team. MJD range indicates the time range of input data used for polynomial fitting. The number given in parentheses represents the 1$\sigma$
uncertainty in the least significant digit of a stated value.}
\begin{tabular}{ l c c c }
& Interval 1 & Interval 2 & Interval 3 \\
\hline
MJD range & 58210.5 - 58277 & 58243 - 58312 & 58280 - 58350 \\
Epoch (MJD) & 58243.0(6) & 58277.02(7) & 58318.99(2) \\
$\nu$ (Hz) & 0.0572491(1) & 0.05737625(8) & 0.0574503(2) \\
d$\nu/$dt \,\,\,\,\,\, (10$^{-11}$ Hz s$^{-1}$) & 5.630(3) & 3.034(3) & 0.950(3) \\
d$^2\nu/$dt$^2$ \, (10$^{-18}$ Hz s$^{-2}$) & -2.51(1) & -6.65(1) & -3.19(1) \\
d$^3\nu/$dt$^3$ \, (10$^{-23}$ Hz s$^{-3}$) & -1.450(1) & 0.3199(9) & 0.5883(8) \\
d$^4\nu/$dt$^4$ \, (10$^{-30}$ Hz s$^{-4}$) & 6.693(1) & -0.450(1) & -0.714(1) \\
d$^5\nu/$dt$^5$ \, (10$^{-36}$ Hz s$^{-5}$) & 8.769(6) & -3.187(1) & -7.849(1) \\
\end{tabular}
\label{timing_params}
\end{table*}
\subsection{Torque--Luminosity Modeling}
\label{Sec:torque-luminosity}
In order to investigate the torque--luminosity correlation of the source, we measure the spin frequency derivative ($\dot{\nu}$) by fitting a linear function to each consecutive three spin frequency measurements.
The uncertainty in $\dot{\nu}$ measurements are calculated from the error range of the slope of the linear function on $1\sigma$ level. X-ray luminosity of 2S 1417$-$624 (at 10 kpc and $D_{10}\equiv d/10$ kpc) is calculated from the count rate history provided by \emph{SWIFT}/BAT team\footnote{\label{swiftfn}\url{https://swift.gsfc.nasa.gov/results/transients/}} \citep{Krimm2013} by multiplying a flux conversion factor of $1.13 \times 10^{-7}$ erg cts$^{-1}$ \citep{Ji2020}.
\begin{figure}
	\includegraphics[width=\columnwidth]{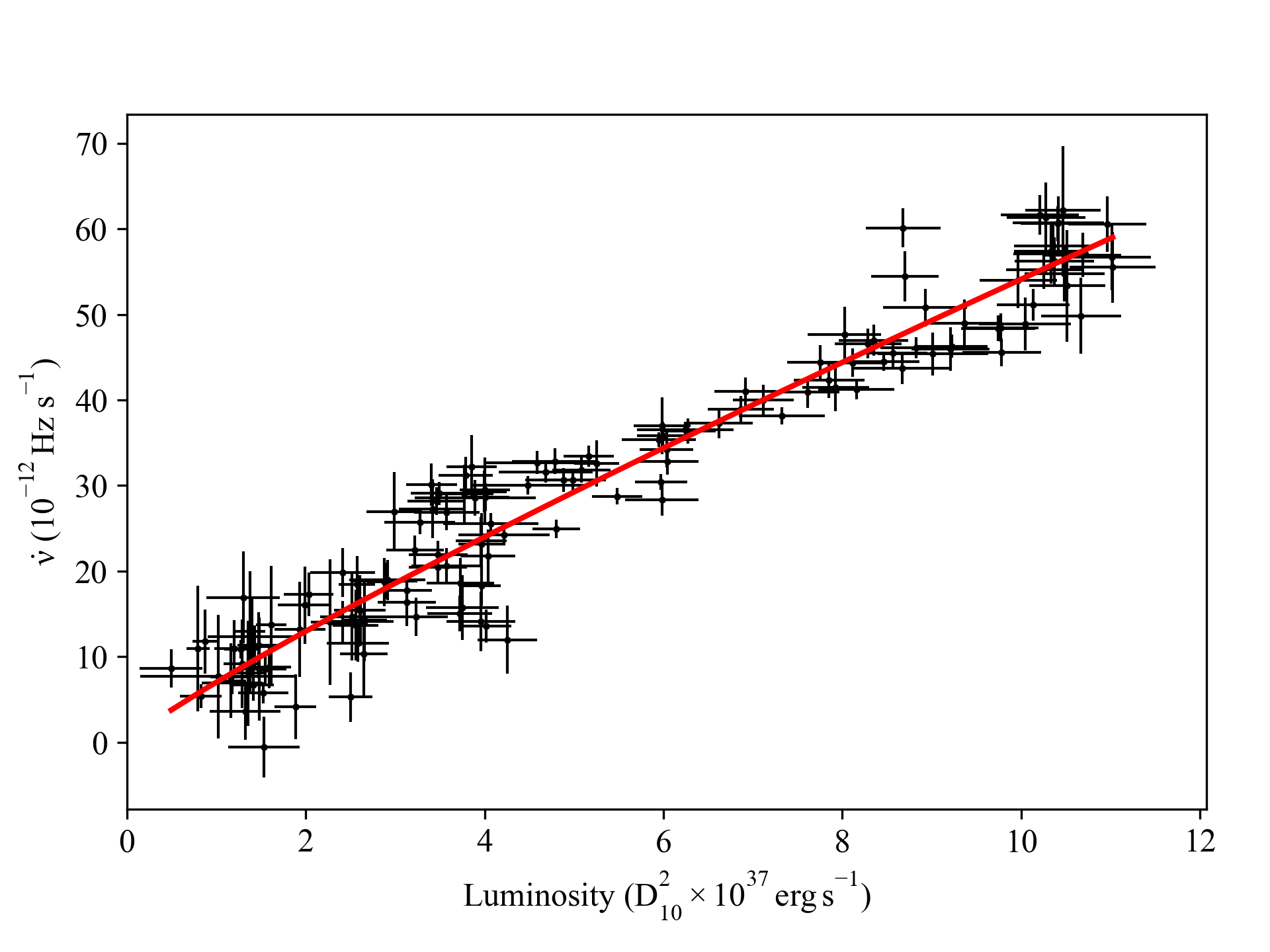}
	\caption{The spin-up rate vs. luminosity relation of 2S 1417$-$624. The solid red line indicates the best fit for the $\dot{\nu}_{\rm{model}}$.}
	\label{fig_torque_L}
\end{figure}
Furthermore, to estimate the X-ray luminosity ($L_{x}$) at times corresponding to that of the $\dot{\nu}$ measurements, we use linear spline interpolation with \texttt{interp1d}\footnote{\url{https://docs.scipy.org/doc/scipy/reference/tutorial/interpolate.html}} tool of \texttt{scipy} library of Python. During the interpolation, we scale the uncertainty in luminosity by 5\% for systematic errors. The spin-up rate vs. $L_x$ variation is shown Fig. \ref{fig_torque_L}. Since most of the disk--magnetosphere interaction models (e.g. \citep{Ghosh1979,Lovelace1995,Kluzniak2007} suggest a relation between X-ray luminosity and the spin-up rate in the form of $\dot{\nu} \propto L_{x}^{\alpha}$, we model the relation as:
\begin{equation}
\dot{\nu}_{\rm{model}} = \beta \, L_{x}^{\alpha}
\end{equation}
where the coefficient $\beta$ and index $\alpha$ are allowed to vary. The fitting procedure yields $\beta =7.0\pm 0.4$ and $\alpha=0.89\pm0.03$. It should be noted that the power-law dependence on luminosity is compatible with the disk accretion model introduced by \citet{Ghosh1979} which envisages $\alpha= 6/7\sim 0.86$. Making use of $\dot{\nu}_{\rm{model}}$ parameters, we generate a model series for the frequency evolution as:
\begin{equation}
\nu_{\rm{model}}(t) = \nu_0 + \int_{t_0}^{t} \dot{\nu}_{\rm{model}}(t') dt'
\end{equation}
where $t_0$ is the time of burst onset and $\nu_0$ is the spin frequency at $t_0$. The numerical integration is carried out using the composite trapezoidal rule. The resultant $\nu_{\rm{model}}(t)$ which overlays the spin frequency measurements is demonstrated in Fig \ref{fig_freq_model}. The spin frequency evolution of 2S 1417$-$624 is adequately described by the constructed model, $\nu_{\rm{model}}(t)$, within the residuals of few $\micro$Hz.
\begin{figure}
	\includegraphics[width=\columnwidth]{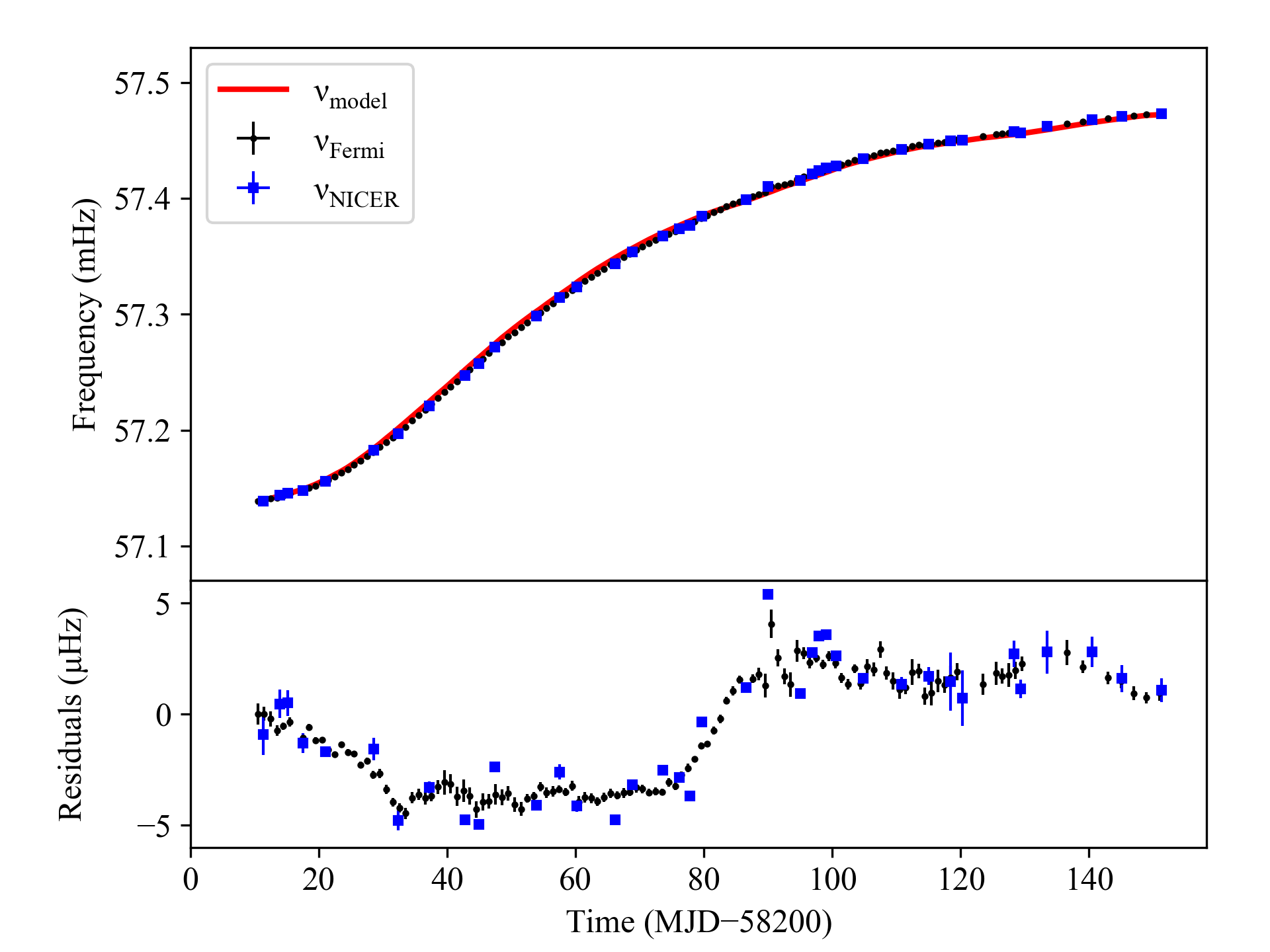}
	\caption{Upper panel: The spin frequency evolution of 2S 1417$-$624 during its outburst in 2018. Blue squares: \emph{NICER}/XTI, black dots: \emph{Fermi}/GBM, solid red line: $\nu_{\rm{model}}(t)$. Bottom panel: Residuals after the removal of $\nu_{\rm{model}}(t)$ from the spin frequency measurements. }
	\label{fig_freq_model}
\end{figure}

\subsection{Noise Strength Analysis}
In order to understand the noise processes that lead to observed pulse frequency fluctuations, we examine the residuals after removing $\nu_{\rm{model}}$ from the spin frequency measurements. The pulse frequency fluctuations are analyzed using the root-mean-square (rms) residual
method developed by \cite{Cordes1980} and \cite{Deeter1984}. In this technique, spin evolution of a pulsar is described as polynomials of degree $m$, and the mean squared values of residuals $\langle\sigma_R^2(m, T)\rangle$ are used to estimate the r$^{th}$-order red noise strength $S_r$ on a timescale $T$. After removing the regular spin-up trend, the noise strength can be estimated as:
\begin{equation}
S_r = T^{2r-1} \dfrac{\langle\sigma_R^2(m, T)\rangle}{\langle\sigma_R^2(m, 1)\rangle_u}
\end{equation}
where $\langle\sigma_R^2(m, 1)\rangle_u$ is the normalization coefficient of unit noise strength. The normalization coefficient can be estimated by measuring the expected mean square of residuals after removal of $m^{th}$ degree polynomial for the unit strength red noise ($S_r=1$) on unit timescale ($T=1$).
\begin{figure*}
	\includegraphics[width=0.49\columnwidth]{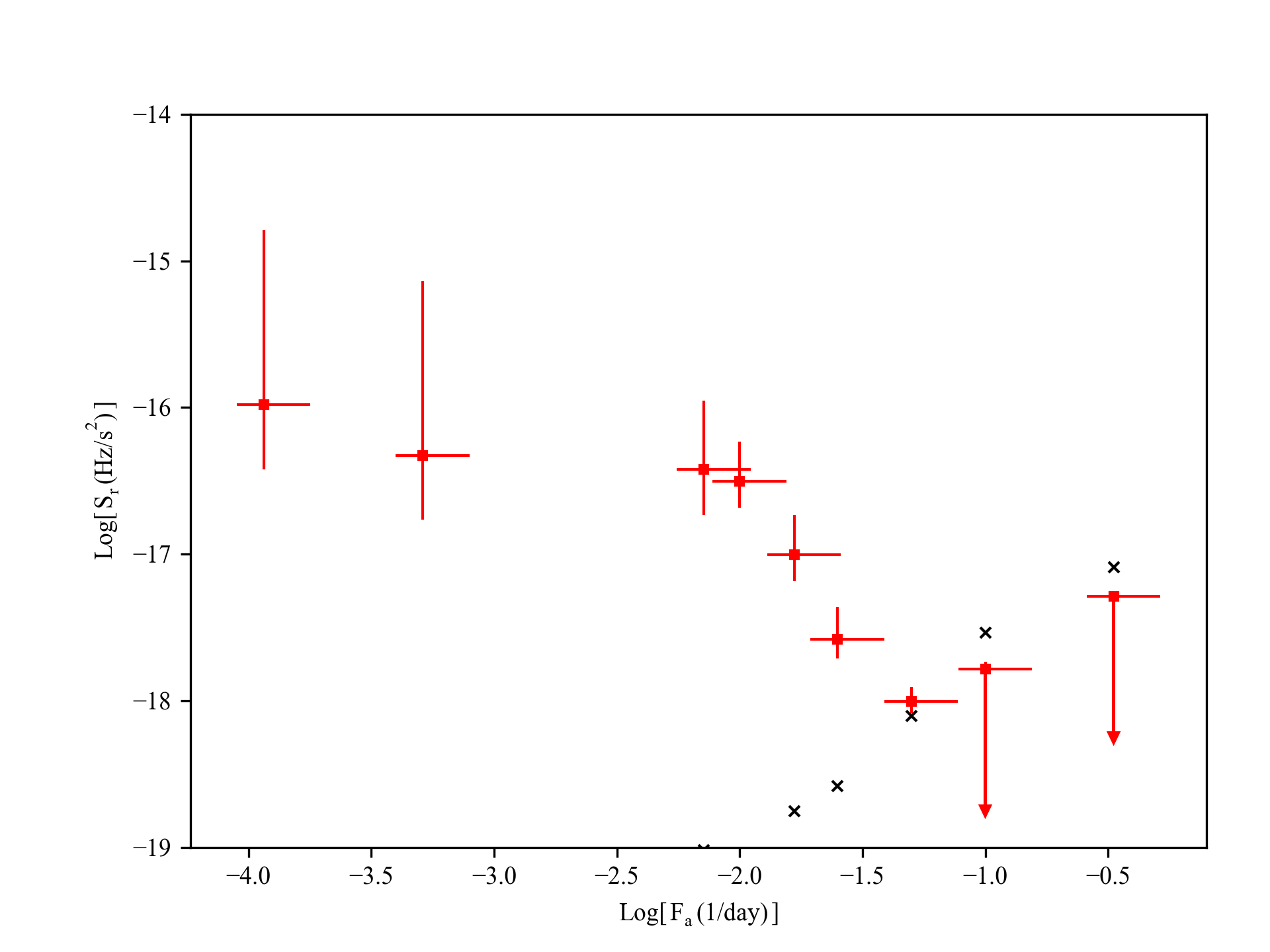}
	\includegraphics[width=0.49\columnwidth]{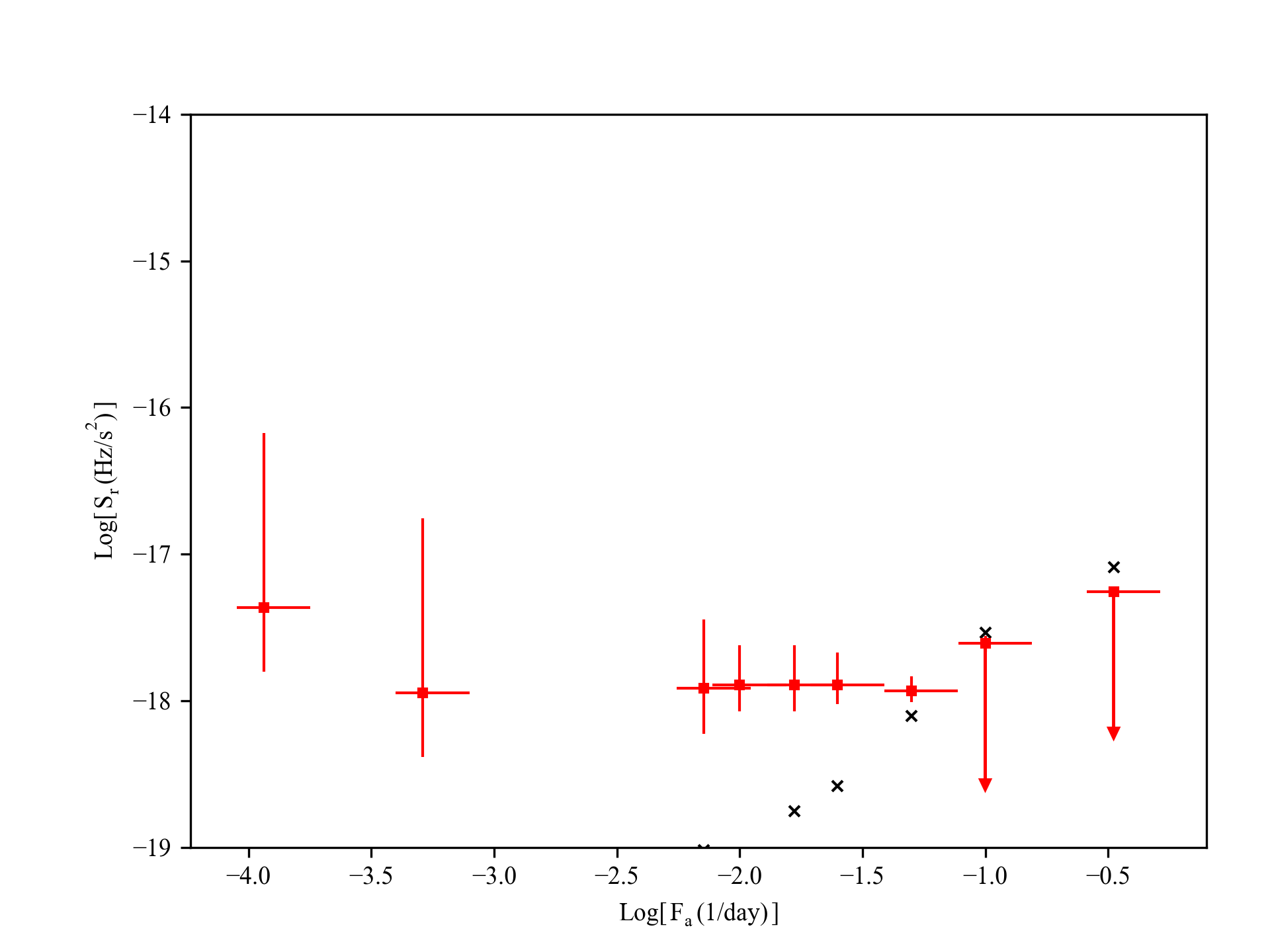}
	\caption{The power density spectra of pulse frequency fluctuations of 2S 1417$-$624, generated from the residuals after removing $\nu_{\rm{model}}$ (right panel) and from the residuals after removing polynomial models (left panel). The black crosses indicate the measuremental noise levels.}
	\label{fig_noise}
\end{figure*}
The normalization coefficients for different orders of red noise $r$ and degree of polynomials $m$ can be calculated via numerical simulations \citep{Scott2003} or direct evaluations \citep{Deeter1984}. While using $\nu_{\rm{model}}$ to describe the spin frequency evolution, we estimate the normalization coefficient of the simulated data series after the removal of $\nu_{\rm{model}}$. 

We calculate the noise strength at various timescales depending on the data sampling rate of the spin frequency measurements. Furthermore, we also include previous spin frequency measurements provided by \emph{BATSE} team\footnote{\label{batsefn}\url{https://gammaray.nsstc.nasa.gov/batse/pulsar/}} in order to extend the maximum timescale of the noise strength estimations and to obtain statistical improvements for power density estimates. Hence, we also perform exact analysis procedure to the \emph{BATSE} measurements as well.
In each timescale, the noise strength measurements are logarithmically averaged and combined into a single power estimate. Finally, by mapping the power density estimates to the corresponding analysis frequencies (F$_a$ = 1/$T$, where $T$ is the timescale in which the noise analysis is performed), we generate the power density spectra of pulse frequency fluctuations (Fig. \ref{fig_noise}). In order to understand the effect of the accretion model on the shape of the power density spectra, we also construct a power density spectrum of the same data set without employing $\nu_{\rm{model}}$. Videlicet, we investigate the residuals after removal of quadratic polynomial trends in the data set ($m=2$) for each timescale and calculate the corresponding power density estimates. When $\nu_{\rm{model}}$ is not employed, we make use of normalization coefficients introduced in Table 1 and 2 of \cite{Deeter1984}.

We find that without employing the $\nu_{\rm{model}}$, the power density spectrum possesses a red noise component with a slope of $\sim$--2 in between the analysis frequencies $ -1.3\gtrsim \textrm{log}(\textrm{F}_a) \gtrsim -2.2$. On longer timescales (i.e. $\textrm{log}(\textrm{F}_a)\lesssim -2.2)$, the spectral shape is flattened out and altered to a white noise structure (Fig. \ref{fig_noise}, left panel). In this case, the noise strength, $S_r$, varies in the range of $\sim$10$^{-18}$ Hz s$^{-2}$ to $\sim$10$^{-16}$ Hz s$^{-2}$. On the other hand, if the $\nu_{\rm{model}}$ is used as a model for describing the rotational evolution of the source, the noise level is reduced approximately 2 orders of magnitude for the timescales in which a significant power density estimate is attained and the power density spectrum results in a flat shape ($S_r \sim10^{-18}$ Hz s$^{-2}$) that is consistent with white noise structures in all timescales (Fig. \ref{fig_noise} right panel). In both cases, the power density spectra are dominated by measuremental noise level for analysis frequencies higher than $-1.3$.

\section{Discussion}

In this study, we examine the characteristics of the BeXB 2S 1417$-$624 through spectral and timing analysis of the \emph{NICER} observations during its outburst in 2018.
In the spectral analysis, we model the individual spectra of each observation with high-energy cut-off and partial covering fraction absorption (PCFA) model to investigate the flux-dependent spectral variations of the source throughout the 2018 outburst for the first time. We observe that photon index exhibits an anti-correlation with 0.8--12 keV flux below $\sim$0.7 $\times$ 10$^{-9}$ erg s$^{-1}$ cm$^{-2}$. Such an anti-correlation was also observed in both 1999 \citep{Inam2004} and 2009 outbursts \citep{Gupta2018}, for which the source was interpreted to be in sub-critical accretion regime. In this regime, the height of the emission region inside the accretion column suggested to be inversely proportional to the X-ray luminosity \citep{Becker2012}. The Comptonisation takes place in the zone, between the radiative shock and neutron star surface. This zone shrinks as the luminosity increases which leads to an increase in the optical depth. Thus, the spectrum gets harder, which eventually yields an anti-correlation between luminosity and photon index \citep{Reig2013}.The 2018 outburst of the source was significantly more luminous than the previous outbursts \citep{Ji2020}, allowing us to examine a broader luminosity range. We find that the relation between the photon index and 0.8--12 keV flux switches from anti-correlation to a slight positive correlation above $\sim$0.7 $\times$ 10$^{-9}$ erg s$^{-1}$ cm$^{-2}$ level (Fig. \ref{fig_specevo_time}). 
The transitional flux level is consistent with the $L_{\rm{crit}}$ estimates that are deduced from variations of the pulse profile morphology and pulse fraction-luminosity correlations \citep{Gupta2018,Ji2020}. Beyond this flux level, the source possibly enters to super-critical accretion regime in which the radiation pressure is high enough to halt the accretion flow at a certain height above the pulsar, thus the beam geometry is altered from pencil beam to fan beam \citep{Becker2012}. In this case, comptonization diminishes since the accelerating seed electrons are reduced due to the shock in the accretion column \citep{Becker2012}, giving rise to the observed positive correlation between photon index and flux. We observe the effect of this transition on the covering fraction and the column density of partially covering absorber n$_{H_2}$ as well. The n$_{H_2}$ values remain uncorrelated below $L_{\rm{crit}}$ and significantly rise above $L_{\rm{crit}}$. Some transient X-ray binaries are shown to have an increased absorption on the phases that are associated with the dips in their pulse profile (e.g. GRO J1008-57: \cite{Naik2011}, GX 304-1: \cite{Jaisawal2016}). \citet{Ji2020} studied the pulse profile evolution of 2S 1417-624 during the outburst in 2018 with the same NICER data set. In their study, they reported that the pulse profiles have complex shape and are luminosity dependent. It is interesting to note that above $L_{\rm{crit}}$, the pulse profiles exhibit another dip, (see Fig.2 of \cite{Ji2020}, around phase 0.6). The increased absorption at the super-critical regime may also be associated with this observed changes in the pulse profile.  On the other hand, it is also possible that n$_{H_2}$ may increase due to the absorption of the material around the neutron star if there is a formation of an envelope around the neutron star that resembles the case of Swift J0243.6+6124 \citep{Zhang2019}.

Similar to the case of Swift J0243+6124 \citep{Doroshenko2020}, \citet{Ji2020} suggests that another transition takes place between gas dominated disc and radiation dominated disc for 2S 1417$-$624 at the luminosity level $L_{\rm{zoneA}} \sim 7\times 10^{37} D_{10}^{2}$ erg s$^{-1}$. At this stage, the radiation pressure becomes so strong that it influences the disc structure and leads to alterations in the emission region and the pulse profile. The transition to $L_{\rm{zoneA}}$ is shown to affect the pulse profile shape (Swift J0243+6124, \citet{Doroshenko2020}; 2S 1417$-$624, \citet{Ji2020}). Furthermore, when studying the hardness ratio of 2--20 keV to 15--50 keV emission (i.e. \emph{MAXI} and \emph{Fermi}/GBM count rates), \citet{Doroshenko2020} noticed that $L_{\rm{zoneA}}$ transition also leads to an increase in the soft emission of Swift J0243+6124. We do not find any significant change in flux dependence of spectral features in super-critical regime and radiation pressure dominated disc regime in our analysis for 2S 1417$-$624 within 0.8--12 keV band. It should be noted that \citet{Wilson2018} investigated the hardness intensity diagrams in both soft ([7--10 keV]/[4--7 keV] bands with \emph{NICER}) and hard energy bands ([12--16 keV]/[8--12 keV] bands with \emph{Fermi}/GBM) for Swift J0243+6124. In the hardness intensity diagrams within the hard band, they noticed a transition which takes place in a higher luminosity level than $L_{\rm{crit}}$ (perhaps due to $L_{\rm{zoneA}}$ transition \citep{Doroshenko2020}), but attributed to different reasons. However, in the hardness intensity diagram obtained with \emph{NICER}, they noticed only the first transition which is taking place at $L_{\rm{crit}}$ level. Thus,
$L_{\rm{zoneA}}$ transition may lead to observable spectral changes in higher energy bands (e.g. as in the case of Swift J0243+6124 \citep{Kong2020})  but the consequences of such a transition are possibly not reflected or resolvable on the spectra below $\sim12$ keV in our study of 2S 1417$-$624. Therefore, observations with X-ray missions which are sensitive to photons in a broader energy range, such as \emph{NuSTAR} or \emph{HXMT}, are required to reveal the changes in the spectral features that are expected for $L_{\rm{zoneA}}$ transition.

\begin{figure}
	\includegraphics[width=\columnwidth]{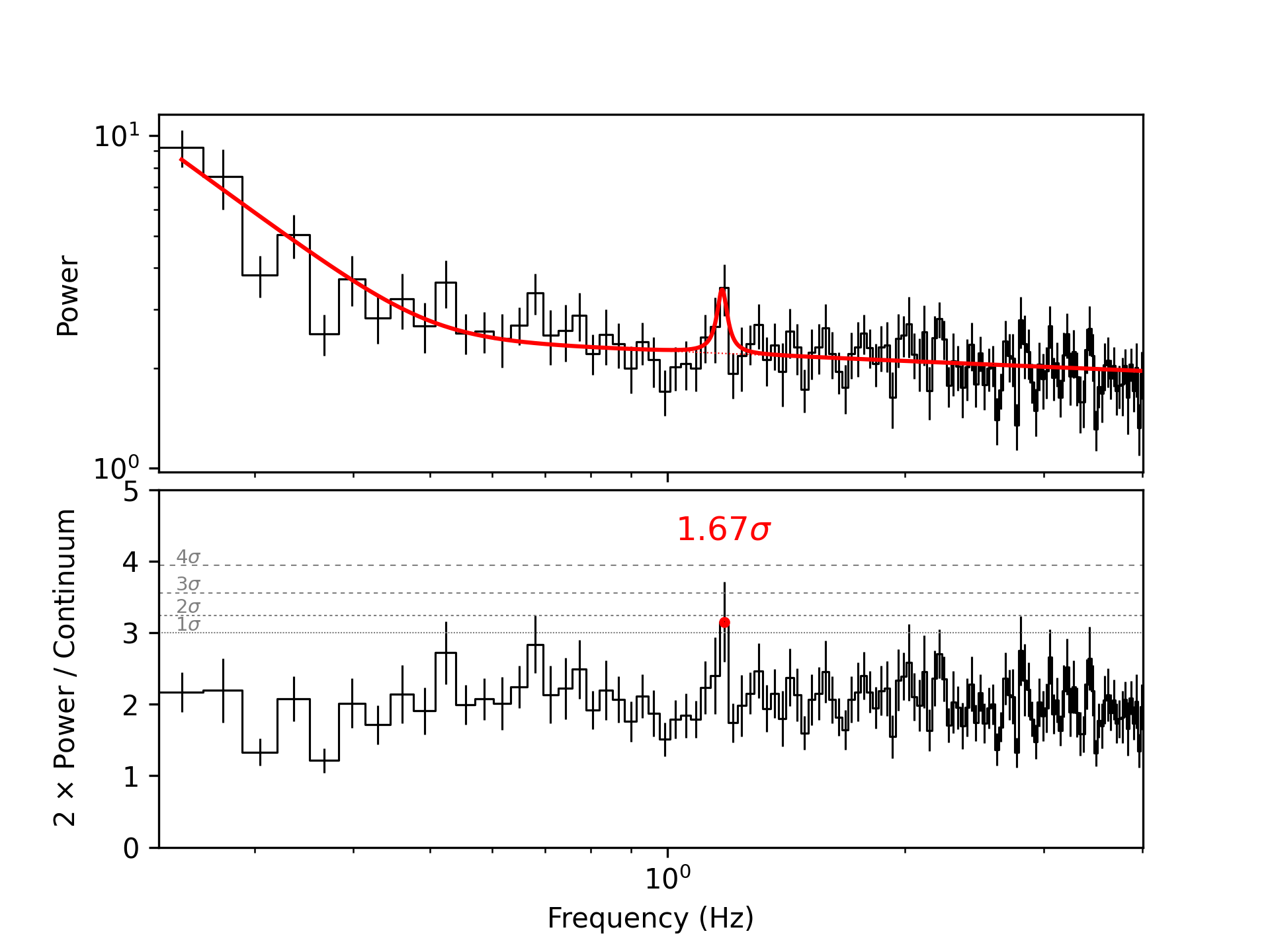}
	\caption{Top panel: The power spectrum (in black) is created by splitting the \emph{NICER}/XTI lightcurve (Obs. ID. 1200130123, taken on MJD 58245) into 64 s time intervals (using 512 bins with 0.125 s rebinning time), averaging these time intervals into one frame (in order to increase the statistical significance of the spectral features \citep{vanderKlis1988}), and rebinning the resulting spectrum by 2. A smooth broken power-law with a Lorentzian component is fitted to the power spectrum (in red). Bottom panel: The power spectrum is normalized by dividing it to the modelled continuum, and the result is multiplied by two. The dashed gray lines indicate 1, 2, 3, and 4 sigma levels of significances. }
	\label{fig:QPO}
\end{figure}

In the timing analysis, we enrich the frequency measurements provided by \emph{Fermi} GAPP team with additional measurements obtained from \emph{NICER} observations. We first use this spin frequency history to model the relation between observed spin-up rates and the luminosity calculated for 10 kpc distance (see Section \ref{Sec:torque-luminosity}). We find that the correlation between these two parameters can be described as $\dot{\nu}_{12} = (7.0\pm 0.4) \times L_{37}^{0.89\pm0.03}$. Provided that the observed X-ray luminosity is proportional to the total bolometric luminosity, this correlation can be interpreted as accretion from a disc when there is a net positive torque on the pulsar \citep{Ghosh1979}. Correlation between the spin-up rate and X-ray luminosity in different X-ray energy bands was observed in this outburst and the previous outbursts of 2S 1417$-$624 (e.g. \cite{Finger1996,Inam2004,Ji2020}), and also in other transient systems (e.g. GRO J1744$-$28, \citep{Bildsten1997}; SAX J2103.5+4545, \citep{Baykal2007}; Swift J0243.6+6124, \citep{Zhang2019}). The power-law index found from modeling the correlation is consistent with the Ghosh \& Lamb picture which predicts the index as $\alpha=6/7$. In this case, the relation between the luminosity and spin-up rate can be described as \citep{Ghosh1979,Sugizaki2017}:
\begin{equation}
\dot{\nu}_{12} = 1.4 \mu_{30}^{2/7} n(\omega_s) R_6^{6/7} M_{1.4}^{-3/7} I_{45}^{-1} L_{37}^{6/7}
\end{equation}
where $\dot{\nu}_{12}$ is the spin-up rate (in $10^{-12}$ Hz s$^{-1}$), $\mu_{30}$ is the magnetic dipole moment (in Gauss cm$^3$), $n(\omega_s)$ is the fastness parameter, $R_6$ is the neutron star radius (in $10^{6}$ cm), $M_{1.4}$ is the mass of the pulsar (in 1.4 $M_{\odot}$), $I_{45}$ is the moment of inertia (in $10^{45}$ gr cm$^{2}$) and $L_{37}$ is the bolometric luminosity (in $10^{37}$ erg s$^{-1}$). For slow rotators, such as 2S 1417$-$624, the fastness parameter can be approximated as $n(\omega_s) \sim 1.4$. If the calculations are proceeded for 10 kpc distance with the typical neutron star parameters such as $ R_6 \sim 1$, $M_{1.4} \sim 1$ and $I_{45} \sim 1$, it leads to a magnetar-like dipole field strength of $\sim$10$^{14}$ G which seems implausible. This issue was also noticed by \citet{Sugizaki2017} and \citet{Ji2020}. While studying the outburst in 2009, \citet{Sugizaki2017} used a correction factor of $\sim$4.5 between the measured spin-up rate and the spin-up rate that is used for the Ghosh \& Lamb model assuming a fixed magnetic field strength of $2.6\times 10^{12}$ G for 2S 1417$-$624. \citet{Ji2020} suggested that either the source is located at twice the distance of the \emph{Gaia} estimate, or the torque models should be modified by quadrupole fields to explain higher torques generated at such luminosity levels. \citet{Long2007} investigated the pure dipole, pure quadrupole and mixed dipole/quadrupole fields and their effects on accretion torques. Even though mixed fields may help explain the complex pulse profiles, as in the case of 2S 1417$-$624, existence of quadrupole fields essentially reduces the net positive torque exerted on to the pulsar \citep{Long2007}. Hence, quadrupole fields seems to be an unlikely explanation for high-accretion torques observed in 2S 1417$-$624. Nevertheless, if the source is located at a greater distance than the \emph{Gaia} estimate (e.g. 20 kpc \citep{Ji2020}), then it is possible to reduce the dipolar field strength of pulsar to canonical values (of the order $10^{12}$ G). One other possibility is that 2S 1417$-$624 may host an unexpectedly low mass neutron star, like SMC X-1 \citep{Kaper2006}, which can at least partially explain the higher accretion torques generated at such luminosity levels. We also seek for quasi-periodic oscillations (QPOs) within the Fourier spectra of the source with an aim of comparing the inferred inner disc radii of the accretion models. We observe low signal-to-ratio features around $\sim$1.2 Hz in one observation right after the peak of the outburst, but unfortunately none of these features are statistically significant enough to be classified as a QPO (e.g. see Fig. \ref{fig:QPO}). The QPOs are generally associated with the inhomogeneities at inner accretion disc. If this tentative feature is assumed as a QPO whose centroid frequency corresponds to the orbital Keplerian frequency $\nu_k$, then the inner disc radius can be estimated via $r_{0} = (\sqrt{GM}/2\pi \nu_k)^{2/3}$ which yields $\sim1.5\times 10^8$ cm. The inner disc radius can also be expressed as $r_0 \sim 0.52 \mu^{4/7} (2GM)^{-1/7} \dot{M}^{-2/7}$ where $\dot{M}$ is the mass accretion rate \citep{Ghosh1979}. When the feature is observed, the source luminosity was $\sim 1\times 10^{38} (D_{10}^{2})$ erg/s. Using the relation $L \simeq GM\dot{M}/R$, this tentative feature implies a dipolar magnetic field strength on the the order $10^{12}$ G. However, we stress that the signal is too weak to reach a definite conclusion. Therefore, an independent magnetic field strength measurement using cyclotron lines is crucial to understand the nature of 2S 1417$-$624 as suggested by \citet{Ji2020} and also continous monitoring campaigns during outburst are also important for revealing QPO features.

Torque fluctuations and the timing noise properties of many isolated pulsars and persistent accretion powered pulsars have been studied in the literature (\cite{Baykal1993,Bildsten1997, DAlessandro1995,Baykal1997,Bildsten1997}). In this study, we use all the spin frequency and flux measurements provided by \emph{BATSE}, \emph{Swift}/BAT and \emph{Fermi}/GBM; and we, for the first time, examine the power density spectrum of the pulse frequency fluctuations of a transient system, namely 2S 1417$-$624, for two different cases. In the first approach, we extracted the power density spectrum using the standard technique \citep{Deeter1984} where the residuals after quadratic polynomial trends are used for noise strength calculations. In the second approach, we construct the power density spectrum for the residuals after describing the frequency evolution with $\nu_{\rm{model}}$, which is obtained from the torque--luminosity relation. The second approach is performed with an aim of understanding the effect of disc accretion torques on the power density spectrum.

In the first case, we find a power density spectrum with a steep red noise component ($\Gamma \sim -2$) in the analysis frequency range of $ -1.3\gtrsim \textrm{log}(\textrm{F}_a) \gtrsim -2.2$, and the power density estimates drifts from $\sim$10$^{-18}$ Hz s$^{-2}$ to $\sim$10$^{-16}$ Hz s$^{-2}$ for lower analysis frequencies. Many persistent systems, which accrete from a disc, are shown to have red noise components in the power density spectrum. For example, Cen X-3 exhibits a red noise component with $\Gamma \sim -1$ and its noise strength varies in the range of $\sim$10$^{-18}$ Hz s$^{-2}$ to $\sim$10$^{-16}$ Hz s$^{-2}$ \citep{Bildsten1997}. SAX J2103.5+4545 shows similar noise strength levels and has the steepest red noise component with $\Gamma \sim -2.13$ among high mass X-ray binaries \citep{Baykal2007}. The noise strength levels and red noise steepness found for 2S 1417$-$624 are consistent with aforementioned disc-fed systems, which confirms the formation of a transient disc on the timescales of the outburst duration. For longer timescales, (i.e. $\textrm{log}(\textrm{F}_a)\lesssim -2.2)$, the power density spectrum is flattened out to $S_r \sim10^{-16}$ Hz s$^{-2}$ level. This high level of timing noise may indicate that the accretion still continues in the quiescent phases between outbursts, possibly due to the cold disc which is depleted after the outbursts as suggested by \citet{Tsygankov2017}.

In the second case, when the $\nu_{\rm{model}}$ is used as a model for describing the rotational evolution of the source, we obtain a flat power density spectrum with $S_r \sim10^{-18}$ Hz s$^{-2}$. The shape of the power spectrum is compatible with white torque-noise structure and it has two important aspects. First, the spin frequency evolution generated from disc models is adequate to describe the observed spin frequencies and conveniently eliminates almost all of the red noise component of the power spectrum that arises from disc accretion. Secondly, assuming that the contribution of the disc is removed from the power density spectrum, the remaining noise level is also in agreement with wind-fed systems. For example, wind-fed systems such as Vela X-1, 4U 1538$-$52 and GX 301$-$2 are also known to exhibit white noise structures and these sources have power spectra with noise strengths in the range of 10$^{-20}$--10$^{-18}$ Hz s$^{-2}$ \citep{Bildsten1997}. This may indicate that 2S 1417$-$624 is also subject to low-level wind material from the companion, apart from the accretion disc which is more dominant and enduring for all timescales. In both power spectra, unfortunately the power density estimates are dominated by measuremental noise levels for analysis frequencies higher than log(F$_{a}$) $\sim$--1.3, which prevents us from examining the timing noise on dynamical timescales. Thus, continuous and more precise spin frequency measurements are required to understand the neutron star's interior response in the case of accretion.

\section*{Acknowledgements} 
We acknowledge support from T\"{U}B\.{I}TAK, the Scientific and Technological Research Council of Turkey through the research project MFAG 118F037. This research has made use of the \emph{Swift}/BAT transient monitor results provided by the \emph{Swift}/BAT team and the \emph{Fermi} data provided by the NSSTC and the GAPP team. We acknowledge the use of public data from the \emph{NICER} archive. We would like to express our gratitudes to the anonymous referee for the valuable remarks that
ease the improvement of the manuscript.

\section*{Data Availability}

The observational data used in this study can be accessed publicly from \emph{NICER} Master Catalog of HEASARC archive. Additionally, \emph{Fermi}/GBM and \emph{BATSE} spin frequency time series, and \emph{Swift}/BAT lightcurves are publicly accessible from their corresponding websites (see Footnotes \ref{gbmfn}, \ref{swiftfn}, \ref{batsefn}).
 



\bibliographystyle{mnras}
\bibliography{bib} 








\bsp	
\label{lastpage}
\end{document}